\documentstyle[preprint,aps,version2,10pt]{revtex}
\begin{document}

\begin{quote}
\raggedleft SU-GP-92-3/1 \vspace{-3mm}\\ hep-th/9203040 \vspace{-3mm}\\
March 13, 1992
\end{quote}
\baselineskip=16 truept
\begin{title}
{\Large \bf  How the Jones Polynomial gives rise to physical  \\
\centerline{states of quantum General Relativity}}
\end{title}
\author{Bernd Br\"ugmann$^{(1)}$, Rodolfo Gambini$^{(2)}$,
Jorge Pullin$^{(3)}$}
\begin{instit}
(1) Physics Department, Syracuse University, Syracuse, NY 13244
\end{instit}
\begin{instit}
(2) Instituto de F\'{\i}sica, Facultad de Ciencias, Tristan Narvaja 1674,
Montevideo, Uruguay
\end{instit}
\begin{instit}
(3) Department of Physics, University of Utah, Salt Lake City, UT 84112
\end{instit}
\begin{abstract}
Solutions to both the diffeomorphism and the hamiltonian constraint of quantum
gravity have been found in the loop representation, which is
based on Ashtekar's new variables. While the diffeomorphism constraint is
easily solved by considering loop functionals which are knot invariants,
there remains the puzzle why several of the known knot invariants are also
solutions to the hamiltonian constraint. We show how the Jones polynomial
gives rise to an infinite set of solutions to all the constraints of quantum
gravity thereby illuminating the structure of the space of solutions
and suggesting the existance of a deep connection between quantum
gravity and knot theory at a dynamical level.
\end{abstract}

\eject

An important step in the canonical quantization of 4-dimensional
general relativity is the construction of the space of physical
states, that is the space of wavefunctions which are annihilated by
the diffeomorphism and hamiltonian constraint operators.  The
discovery of the Ashtekar variables for canonical 3+1 dimensional
general relativity \cite{As} has led to the construction of a loop
representation for quantum gravity, in which wavefunctions are
functionals of loops \cite{RoSm}. In this context the diffeomorphism
invariance is quite elegantly described by knot theory. Since knot
invariants are diffeomorphism invariant loop functionals, one is
naturally led to consider knot invariants as candidates for the states
of quantum gravity.

This approach is reminiscent of the canonical quantization in the
geometrodynamical variables, where the diffeomorphism constraint is
formally solved by choosing the states to be functionals of
three-geometries. The hard part is to solve the hamiltonian constraint
(the Wheeler-Dewitt equation).  Since this constraint encodes the
dynamical evolution of the Einstein equations one does not expect that
notions from knot theory are going to be helpful for finding
solutions.  The main intention of this essay is to show that, quite
surprisingly, there exists a connection between knot theory and
quantum gravity {\it also at the dynamical level.} To be more
specific, we will exhibit a solution to all the constraints of quantum
gravity with a cosmological constant in the loop representation
related to the Jones polynomial, and we will indicate how this one
solution gives rise to an infinite new set of vacuum solutions.

It was only quite recently that techniques for concrete calculation of
the action of the constraints in the loop representation were
introduced.  Firstly, through the introduction of a set of
``coordinates'' on loop space \cite{DiGaGrLe} and the advances in the
understanding of Chern-Simons theories \cite{Wi,GuMaMi}, it was
possible to give an explicit, analytic form for several knot
invariants. Secondly, the understanding of the action of the
constraints of quantum gravity in the loop representation has been
greatly enhanced when they were expressed in terms of differential
operators in loop space \cite{Ga,BrPu92}. It was through these
developments that we were in a position to explicitly apply the
diffeomorphism and hamiltonian constraints in the loop representation
to some of the known analytic knot invariants in search for solutions
with triple self-intersections. (A general argument shows that only in
this case one can obtain a nondegenerate metric \cite{BrPu}.)

A positive result was achieved some time ago, when we showed that the
second coefficient of the Alexander-Conway knot polynomial (a
knot invariant associated with the classic Arf and Casson
invariants) was annihilated by all the constraints of quantum gravity
even when considered on knots with triple self-intersections
\cite{BrGaPuprl}. In spite of the appeal of this result, which pointed
out a remarkable connection between gravity and knot theory at a
dynamical level and for the first time exhibited a concrete physical
state of the quantum gravitational field, it was also true that this
one example did not shed much light on the structure of the space of
solutions. The rest of this essay is devoted to show that this
puzzling result emerges quite naturally from simple notions of knot
theory.

The new canonical variables for general relativity introduced by
Ashtekar are a Lie-algebra valued connection $A_a(x)$ and a triad
$E^{b}(y)$ on a three-manifold $\Sigma$, while the three-metric on
$\Sigma$ becomes a derived quantity, $q^{ab}=\mbox{Tr}E^{a}E^b$.
There are two representations of quantum gravity based on the new
variables: the connection representation in which wavefunctions are
functionals of the Ashtekar connection \cite{JaSm}, and the loop
representation in which wavefunctions are functionals of loops
\cite{RoSm}. At a heuristic level, these two representations can be
related by a ``loop transform'' (see below) which maps states and
operators in the connection representation to the loop representation,
in analogy to the Fourier transform which in quantum mechanics relates
the position representation to the momentum representation. Loop
representations have been used for several theories, including Maxwell
electrodynamics \cite{GaTr80,AsRo}, 2+1 gravity \cite{AsHuRoSaSm}, and
even for Yang Mills calculations on the continuum \cite{GaTr} and the
lattice \cite{Br,Fa}. Far from being a mathematical nicety, they are a
quite powerful and concrete way of analyzing the quantum dynamics of
theories based on a connection.

Let us recall that in the connection representation of quantum gravity
based on Ashtekar variables there exists a state that is a solution to
all the constraints given by \cite{Ko,BrGaPunpb},
\begin{equation}
\Psi_{\Lambda}[A] = \mbox{exp} (-\frac{6}{\Lambda}
\int_\Sigma\tilde{\eta}^{abc} \mbox{Tr}[A_a
\partial_b A_c + \frac{2}{3} A_a A_b A_c]).
\end{equation}
That is, the exponential of the Chern-Simons form constructed from the
Ashtekar connection is a solution to all the constraints of quantum
gravity with a cosmological constant $\Lambda\neq0$ in terms of the
Ashtekar new variables. Moreover, the determinant of the three metric
has a definite nonzero value on this state.

Having such a state in the connection representation it is natural to
ask the question if it has a counterpart in the loop representation.
Since the loop transform
\begin{equation}
\Psi[\gamma] =
\int\!{\cal D}A\;\;\mbox{Tr(Pexp}\oint\dot\gamma^a A_a)\Psi[A]
\label{transform}
\end{equation}
is just a formal entity at present (we do not know how to perform the
integral on the right) one simply does not know in general how to find
the counterpart in the loop representation for a given state. However
for the particular state $\Psi_{\Lambda }[A] $, the loop transform
turns out to be identical to the expression for the expectation value
of the holonomy in a Chern-Simons theory.  This can readily be seen by
replacing the expression for $\Psi[A]$ in (\ref{transform}) with
$\Psi_{\Lambda }[A] $. It turns out that this expression has been
evaluated by various techniques in the context of Chern-Simons
theories \cite{Wi,GuMaMi}! The transform $\Psi_{\Lambda}[\gamma]$ is a
knot polynomial which is closely related to the Jones polynomial.
For the particular case considered
$\Psi_{\Lambda}[\gamma]$ is a polynomial in $\Lambda$ where each
coefficient is a knot invariant depending on $\gamma$.  This result
can be generalized to the case of intersecting loops \cite{BrGaPunpb}.

One can therefore conclude that --- formally by construction ---
$\Psi_{\Lambda}[\gamma]$ is a solution to all the
constraints of quantum gravity in the loop representation with a
cosmological constant, and it is nondegenerate if one considers
intersecting loops.  A relevant question is therefore: since we have
the appropriate technology to apply the constraints of quantum gravity
in the loop representation to concrete knot invariants, can we
actually show that this polynomial is a solution to the constraints?
The answer is yes and the result is surprising.

Since we are dealing with knot invariants, we will not be concerned
with the diffeomorphism constraint \cite{diffeo}. The non-trivial
constraint to satisfy is the hamiltonian.  With a cosmological
constant it can be written as
\begin{equation}
{\cal H}_{\Lambda}={\cal H}_{0}+\Lambda\mbox{det}q
\end{equation}
where ${\cal H}_{0}$ is the vacuum hamiltonian constraint and
$\mbox{det} q$ is the determinant of the three metric. We now write
$\Psi_{\Lambda}[\gamma]$ explicitly as a polynomial in $\Lambda$:
\begin{equation}
\Psi_{\Lambda}[\gamma]= c_{0}[\gamma] + c_{1}[\gamma] \Lambda +
c_{2}[\gamma] \Lambda^2 +\ldots{}
\end{equation}
The coefficients $c_{i}[\gamma]$ correspond to concrete analytical
expressions.  For instance, $c_{1}[\gamma]$ is given by the celebrated
expression of Gauss for the self-linking number of $\gamma$,
\begin{equation}
c_{1}[\gamma]=\oint\oint dsdt\;\dot\gamma^{a}(s)\dot\gamma^{b}(t)
\epsilon_{abc} {\gamma^{c}(s)-\gamma^{c}(t)\over |\gamma(s) -\gamma(t)|}.
\end{equation}
This expression is finite despite appearance \cite{Ca}.
$c_{0}[\gamma]$ is 1 when the loop has one connected component and 0 else.
$c_{2}[\gamma]$ can be decomposed as
$c_{2}[\gamma] = c_{1}[\gamma]^{2}+\rho[\gamma]$,
where $\rho[\gamma]$ is a well known knot invariant,
the second coefficient of the Conway polynomial, also related to the
Arf and Casson invariants \cite{GuMaMi}.

Applying the hamiltonian constraint operator based on \cite{Ga,BrPu92}
to the polynomial we obtain again a polynomial in $\Lambda$ each of
whose coefficients should vanish independently. The fact that the
hamiltonian constraint has a homogeneous term and a term linear in
$\Lambda$ means that different orders in the coefficients of
$\Psi_{\Lambda}[\gamma]$ will get mixed. As a final result of the
calculation it turns out that several terms combine to cancel provided
some conditions are fulfilled. These conditions turn out precisely to
require that some portions of the coefficients be annihilated by the
vacuum constraint.  The first one is
\begin{equation}
\hat{\cal H}\, c_{0}[\gamma] =0,
\end{equation}
which is immediate to prove \cite{vanish}. More important is that one finds
\begin{equation}
\hat{\cal H}\,\rho[\gamma] =0.
\end{equation}
That is, $\rho[\gamma]$, the second coefficient of the Conway
polynomial has to be annihilated by the vacuum hamiltonian constraint!
Moreover, it is not annihilated by the determinant of the three
metric, and actually this was crucial for the cancellation of other
terms in the computation.  Hence we have arrived in a simple and
conceptual way at the result first introduced in \cite{BrGaPuprl},
where it was obtained through a laborious computation, that the
second coefficient of the Conway polynomial, a knot invariant, is a
nondegenerate solution to the hamiltonian constraint of quantum gravity.

In fact one can consider the above calculation for higher order
coefficients thereby systematically exploring the structure of these
states, and similar results can be obtained: for each order in the
polynomial there appear portions of the coefficients that are
annihilated by the vacuum hamiltonian constraint and not by the
determinant of the metric. The Jones polynomial therefore emerges as
an infinite tower of physical states of vacuum quantum gravity. This
suggests that a deep and beautiful connection exists between quantum
gravity and knot theory at a level that transcends pure diffeomorphism
invariance and takes into account the full dynamics of general
relativity.

\vskip 3cm
We wish to especially thank Abhay Ashtekar and Lee Smolin for many
fruitful discussions. R.G. also thanks Abhay Ashtekar, Lee Smolin,
Karel Kucha\v{r} and Richard Price for hospitality and financial
support during his visit to Syracuse University and The University of
Utah. This work was supported in part by grant NSF PHY 89 07939 and
NSF PHY 90 16733 and by research funds provided by the Universities of
Syracuse and Utah. Financial support was also provided by CONICYT,
Uruguay.
\vfil


\begin{thebibliography}{99}
\bibitem{diffeo} It should be pointed out that one can
explicitly check that these states are annihilated by the
diffeomorphism constraint in the loop representation, as is done in
\cite{BrGaPuprl}. These manipulations, however, are at present only of
formal character since there are delicate regularization issues that
have to be analyzed in more detail.
\bibitem{vanish} The area derivative that appears in the hamiltonian
constraint annihilates $c_{0}[\gamma]$ since the value of $c_{0}[\gamma]$
is unchanged when one appends an infinitesimal loop.
\end{thebibliography}
\end{document}